\documentclass[preprint,showpacs,preprintnumbers,amsmath,amssymb]{revtex4}

\usepackage{graphicx}
\usepackage{dcolumn}
\usepackage{bm}
\usepackage[dvips]{color}

\begin{document}

\title{Quantum Monte Carlo calculated potential energy curve for the helium dimer}

\author{Xuebin Wu$^{1}$}
\author{Xianru Hu}
\author{Chenlei Du$^{1}$}
\author{Yunchuan Dai$^{1}$}
\author{Shibin Chu$^{1}$}
\author{Leibo Hu$^{1}$}
\author{Jianbo Deng$^{1}$}
\email{dengjb@lzu.edu.cn}
\author{Yuanping Feng$^{2}$}%
 \affiliation{ $^1$Institute of Theoretical Physics, Lanzhou University, Lanzhou 730000, People's Republic of China\\
$^2$Department of Physics, National University of Singapore, 2 Science Drive 3, 117542, Singapore}%

\date{\today}

\begin{abstract}
We report results of both Diffusion Quantum Monte Carlo(DMC) method
and Reptation Quantum Monte Carlo(RMC) method on the potential
energy curve of the helium dimer. We show that it is possible to
obtain a highly accurate description of the helium dimer. An
improved stochastic reconfiguration technique is employed to
optimize the many-body wave function, which is the starting point
for highly accurate simulations based on the Diffusion Quantum Monte
Carlo(DMC) and Reptation Quantum Monte Carlo (RMC) methods. We find
that the results of these methods are in excellent agreement with
the best theoretical results at short range, especially recently
developed Reptation Quantum Monte Carlo(RMC) method, yield
practically accurate results with reduced statistical error, which
gives very excellent agreement across the whole potential. For the
equilibrium internuclear distance of 5.6 bohr, the calculated
electronic energy with Reptation Quantum Monte Carlo(RMC) method is
5.807483599$\pm$0.000000015 hartrees and the corresponding well
depth is -11.003$\pm$0.005 K.
\end{abstract}



\maketitle

\section{Introduction}\label{intro}

The interaction between nobel gas has been a subject of intense
theoretical and experimental studies in the last two
decades\cite{Springall,anderson,heh,he3a,he3b}. Highly accurate
calculations are required to describe the van der Waals (vdW) long
range attraction due to the very small energy differences. Despite
of its simplicity, so far there is no general unanimity about its
equilibrium properties from the experimental side. Actually, it is
very difficult to determine experimentally the complete energy
dispersion. On the other hand, this compound represents a numerical
challenge for theoretical methods, because the general density
approximation (GGA) and other standard treatments based on the
density functional theory (DFT) are not supposed to work well, when
dispersive forces are the critical constituent in the chemical bond.
Despite some progress has been made very
recently\cite{X3LYP,vdw-DFT,VDWDFT,VDW DFT,B2PLYP,DFT-VDW}, a
general and practical answer of this problem is still lacking in the
DFT formalism.

Another family of methods, the sophisticated wavefunction based post
Hartee-Fock methods, such as configuration interaction, coupled
cluster and many-body perturbation theory, could be very accurate
only when they employed extensive large basis sets. However, they
are highly basis-dependent and computational prohibitive, they have
rather steep computational scaling with system size [e.g.,
O($N^{7}$) for CCSD(T), CC with single and double excitations and
perturbational triplets, where M is the number of basis functions],
higher oder coupled cluster (CC) methods are much more expensive.

Quantum Monte Carlo (QMC) methods offer a very promising unified
approach with the accuracy required to van der Waals (vdW) bonded
systems, they are able to deal with a highly correlated variational
wave function, which can explicitly contain all the key ingredients
of the physical system. unlike the sophisticated wavefunction
methods, Quantum Monte Carlo (QMC) methods are basis-free(complete
basis sets) and low scale (often $N^{3-4}$), depending on the
method, which makes the QMC framework generally faster than the most
accurate wavefunction based post Hartree-Fock (HF) schemes for large
enough N. Moreover, recent important developments in the QMC field
allow now to optimize the variational ansatz with much more
parameters and higher accuracy. In turn this can be substantially
improved by projection QMC methods such as the diffusion Quantum
Monte Carlo (DMC)\cite{Foulkes} and Reptation Quantum Monte Carlo
(RMC)\cite{RQMC,pierleoni,Foulkes}.The famous Diffusion Quantum
Monte Carlo method (DMC) have been successfully applied, and highly
accurate results have been obtained for
$He_{2}$\cite{Springall,anderson}, $He-LiH$\cite{anderson},
$HeH$\cite{heh}, $He_{3}$\cite{he3a,he3b}, $H-PsH$\cite{hpsh}.

In the present work we report a systematic study of the helium
dimer, using the latest developments in the QMC framework: an
improved optimization algorithm based on the stochastic
reconfiguration(SR)\cite{SR}technique, the DMC \cite{Foulkes} and
RMC methods\cite{RQMC,pierleoni,Foulkes}. All our QMC calculations
have been performed with the QWalk program \cite{qwalk}.

\section{Method}\label{method}

\subsection{Quantum Monte Carlo}
We use the Variational, Diffusion, and Reptation flavors of Quantum
Monte Carlo (VMC, DMC, and RMC) in our calculations as implemented
in the QWalk program \cite{qwalk}. QWalk use a Diffusion Quantum
Monte Carlo algorithm very similar to that described in Ref
\cite{Foulkes}. The Reptation Quantum Monte Carlo algorithm is from
Ref \cite{pierleoni}, except that we use the approximation to the
Green's function as described in Ref \cite{Foulkes}. These
techniques are able in principle to predict the ground state energy
of the system, since they are based on a direct stochastic solution
of the Schrodinger equation.

However, both the Reptation Quantum Monte Carlo and Diffusion
Quantum Monte Carlo suffer from the sign problem for
fermions\cite{FermionSign}, which forces us to make the fixed-node
approximation, where the nodal surface of the exact wave function is
assumed to be the same as the trial wave function, so the
calculation is highly depend on the trial wave function. This
approximation typically results in recovering 95-99\% of the
correlation energy for single reference system. This the only
uncontrolled approximation in our calculations. All calculations
will be done using this approximation. For more techniques and
implement details, please read
Ref\cite{qwalk,pierleoni,Foulkes,VMCDMC}.

We can control the fixed node approximation\cite{fixnode} somewhat
by varying (optimizing) the orbitals in the trial wave function
using the stochastic reconfiguration(SR)\cite{SR}technique at VMC
level then minimize the DMC and RMC energy. The DMC and RMC
calculations in the HF case were done after optimizing the two-body
and three-body Jastrow factors, without changing the HF determinant.

\subsection{Computational Details}

The trial wave functions of the Slater-Jastrow type,which write as:
\begin{equation}
\Psi_{T}=e^{J}D_{\uparrow}D_{\downarrow}
\end{equation}
where $D_{\uparrow}$ or $D_{\downarrow}$ are Slater determinants of
spin-up(down) electrons and J is a Jastrow factor. This form of wave
function is very general and has been successfully applied to a wide
variety of electronic systems, from single atom to solids containing
many hundreds of electrons\cite{Foulkes}. The single determinant
trial wave functions were constructed using HF orbitals generated
using GAMESS(US) code\cite{gamess} which were then cusp corrected
using the scheme of Ma et al.\cite{cusp} The basis set used a very
large and accurate Atomic Nature Orbits(ANO) style Gaussian
functions on each of the helium atoms, this basis contians
19s9p8d/8s7p6d contacted functions. The exponents of these basis
functions were optimized to provide the lowest HF energy of the
dimer at 5.6 a.u. The Jastrow factor used in this work contains
electron-electron, electron-ion, and electron-electron-ion
correlation terms expanded in natural polynomials with explicit
range cutoffs,this means we use up to Three-Body-Jastrow.

The FN-DMC calculations in this study were performed with a target
population of 5000 configurations, to minimize any population
control error. All calculations contained a minimum of $10^{7}$
statistical samples with extra calculations being performed until
the desired error was arrived upon. This was generally the case for
calculations using small values of the time step, to account for the
increase in serial correlation. To calculate the binding energy in
FN-DMC, we first extrapolated the dimer energies to \emph{timestep}
$\tau=0$ separately. The extrapolation to \emph{timestep} $\tau=0$
was performed by fitting a quadratic polynomial to the DMC data.

While the FN-RMC calculations were also based on the above VMC
optimized trial wave functions. Similar to the FN-DMC calculations,
these calculations contained a minimum of $10^{7}$ statistical
samples with extra calculations being performed until the desired
error was arrived upon. In RMC, the random variable is the path,
which, for finite length, can be a probability density. Therefore,
there's no branching and no population control bias\cite{RQMC}. To
calculate the binding energy in FN-RMC, we extrapolated dimer
energies to \emph{timestep} $\tau=0$ and \emph{length} $N=\infty$
separately. The extrapolation to \emph{timestep} $\tau=0$ and
\emph{length} $N=\infty$ was also performed by fitting a quadratic
polynomial to the RMC data.

\section{Results and Discussion}

Since there is no node for the $1^{1}S$ ground state of helium atom,
QMC simulations can solve this state exactly. Our FN-DMC and FN-RMC
gives the same value with the exact value. The energy calculated for
the dimer at 5.6 bohr for FN-RMC, is 5.807483599$\pm$0.000000015
hartrees, this corresponds to -11.003$\pm$0.005 K; while for FN-DMC,
is -5.807483624$\pm$0.000000029 hartrees, and the computed
interaction energy is -0.000034870$\pm$0.000000029 hartrees, this
corresponds to -11.014$\pm$0.015 K. These results are significant
better results, and they are also in accurate agreement with the
most recent ab initio estimates of the same quantity, namely -10.96
K\cite{Springall}, -10.98 K\cite{EQMC1} -11.008 K\cite{CCFCI3},
-10.998 K\cite{EQMC2}, -11.009 K\cite{cencek}, -10.978 K\cite{ECG},
-10.980 K\cite{ACPF}, $-11.003_{7}$ K\cite{CCSDTQ}, -11.000
K\cite{sapt2}, and 10.9985 K\cite{cencek2}.

Recently, several points of the helium dimer potential were
calculated by Anderson\cite{EQMC2} using the exact (flexible nodes)
quantum Monte Carlo (QMC) method. These results, for the three
distances where a comparison is possible, are consistent with ours
with the error bars overlap. At R=5.6 bohr, Anderson¡¯s result of
-10.998$\pm$0.005 K has only slightly narrow error bar than ours
FN-DMC result, and the same error bar with our FN-RMC result. The
agreement with the 2001 Anderson¡¯s results\cite{EQMC1} is very
good. The present interaction energy at R=5.6 bohr is in good
agreement with the SAPT result from ref\cite{sapt2}, equal to
-11.000$\pm$0.011 K.

A detailed comparison with other high-level ab initio calculated
interaction energies for the helium dimer was presented in
ref\cite{cencek}. Therefore, in the present work we focus on some
new results. For the equilibrium distance, R=5.6 bohr, there exists
now an very interesting improved upper bound to the interaction
energy, -10.9985 K,\cite{cencek2} obtained from four-electron
"monomer contracted" ECG calculations. Our present predictions,
-11.014$\pm$0.015 K and -11.003$\pm$0.005 K, are fully consistent
with this upper bound. Let us further note that for all the
distances considered here, the upper bounds given by
Komasa\cite{ECG} are above our error bars.

In recent, Springall and co-workers reported Quantum Monte Carlo
calculations on the potential energy curve of the helium dimer.
Their FN-DMC result for helium-helium interaction energy at R=5.6
a.u is very accurate, predicted -10.89$\pm$0.17 K or -10.96$\pm$0.15
K, which are in very good agreement with the current accepted values
of around 11.00 K. But their error bars are larger than our FN-DMC
and FN-RMC results. For separated distance at R=4.5 a.u, our FN-DMC
and FN-RMC result are 58.425$\pm$0.034 K and 58.475$\pm$0.013 K,
respectively. And their FN-DMC \cite{Springall} result is
49.76$\pm$0.29 K, while the "exact" QMC 58.3$\pm$0.5 K \cite{EQMC1}.
The well published results are listed in TABLE
I\cite{GQMC,EQMC1,prl,ACPF,CCFCI,SAPT,sapt2,CCSDTQ,ccr12,ECG,ccsdt}.
And for the separated distance at R=4.3 a.u, the corresponding well
published and accepted results are all around 118.0
K\cite{ECCC,1,2,3,4,5}, our FN-DMC and FN-RMC result are
117.879$\pm$0.059 K and 118.019$\pm$0.022 K, respectively. And their
FN-DMC result is 102.24$\pm$0.32 K \cite{Springall}.

As is shown in the TABLE II, our results of FN-DMC and FN-RMC are
totally significant different from the very recent FN-DMC
results\cite{Springall} at some separated distances. For the
separated distance small than R=5.6 a.u, their FN-DMC results are
unambiguous smaller than our FN-DMC and FN-RMC corresponding
results; while for the separated distance large than R=5.6 a.u, like
R=5.9 a.u, their FN-DMC result is -10.31$\pm$0.15 K, and our's
corresponding FN-DMC and FN-RMC results are -10.195$\pm$0.014 K and
-10.184$\pm$0.005 K, respectively. And for R=7.4 a.u, their FN-DMC
and our FN-RMC results are -2.57$\pm$0.18 K and -3.332$\pm$0.002 K,
respectively. As also may be seen in Table II, our FN-DMC and FN-RMC
results are accurate consistent with the SAPT calculations in
ref\cite{sapt2} to fit an analytic potential for the helium
dimer\cite{sapt2} and Hurly and Mehl's \cite{Hurly} analytic
potential. Our FN-RMC results (interaction energies and error bars)
for helium dimer potential energy curve are better.

\section{Conclusions}\label{conclusions}

In summary, in this work we used the FN-DMC and FN-RMC methods to
compute accurate values for the interaction energies of He-He dimer.
Our results show that fixed node DMC and RMC combine with
conventional Slater-Jastrow wave functions can give very accurate
results for a helium dimer, which are of comparable accuracy to many
other methods. The computed values are in excellent agreement with
those of prior "state of the art" electronic structure calculations
available for this system. Due to the aforementioned features, the
relatively new Reptation Quantum Monte Carlo method (FN-RMC) can
easily account much more correlation energy and can systematically
reduce the statistical biases compared to the sophisticated FN-DMC
method \cite{RQMC}, which opens a new path of research for Quantum
Monte Carlo.

As is also shown in present work, the FN-RMC shows generally better
results (more accurate interaction energies and much smaller error
bars) than FN-DMC under the same conditions. Furthermore, the FN-RMC
method have a very similar scaling compare to FN-DMC method, so it
should be a very promising method in the future. Our results also
suggest that the fixed-node error could be almost negligible in the
FN-DMC and FN-RMC calculations for these systems, which has been
suggested previously by Massimo Mella and James B.
Anderson.\cite{anderson}

\begin{acknowledgments}
This work was supported by the fund of State-based Class. We wish to
thank this support. Thanks for the programme packages of Qwalk and
Gamess(US), because this work was based on both of them.
\end{acknowledgments}

\begin{table}[thb]
\begin{ruledtabular}
\begin{center}
\caption{Comparison of selected predictions of the helium-helium
interaction at R=4.5 bohr. Energies in Kelvin.}\label{table.1}
\renewcommand{\thefootnote}{\thempfootnote}
\begin{tabular}{cc}
 Method & Interaction energy
\\ \hline Fixed node DMC\cite{Springall}     & 49.76$\pm$0.29
\\ EQMC \cite{EQMC1}                         & 58.3
\\ ECG \cite{ECG}                            & 58.517
\\ r12-MR-ACPF \cite{ACPF}                   & 58.49
\\ CCSDT(Q)/CBS\cite{CCSDTQ}                 & 58.397
\\ CCSD(T)\cite{sapt2}                       & 59.470
\\ SAPT \cite{sapt2}                         & 58.371
\\        Green function QMC\cite{GQMC}      & 60.0
\\ Analytical potential\cite{prl}            & 60.44
\\ CCSD(T)+FCI correction\cite{CCFCI}        & 59.54
\\  SAPT \cite{SAPT}                         & 58.037
\\ CCSD(T)-R12 \cite{ccr12}                  & 59.543
\\ CCSD(T)/CBS+FCI \cite{ccsdt}              & 58.407

\end{tabular}
\end{center}
\end{ruledtabular}
\end{table}

\begin{table}
\caption{Interaction energy values of the helium dimer at different
values of the atom separation. The interaction energies are given in
K.}
\begin{ruledtabular}
\begin{tabular}{lccccl}
\multicolumn{1}{c}{R(bohr)} & \multicolumn{1}{c}{FN-RMC
\footnote{this work}} & \multicolumn{1}{c}{FN-DMC \footnote{this
work}} & \multicolumn{1}{c}{FN-DMC
\footnote{Reference\cite{Springall}}}&
\multicolumn{1}{c}{Hurly\footnote{Reference\cite{Hurly}}} &
\multicolumn{1}{c}{SAPT FIT\footnote{Reference\cite{sapt2}}}
\\
\hline
 0.9 & 350482.26$\pm$1.44 & 350133.52$\pm$3.64 & 314409.15$\pm$3.79 &   348590.96     & 350152.17   \\
 1.9 & 44852.74$\pm$0.91  & 44801.15$\pm$2.28  &  45959.52$\pm$4.11 &    44863.43     &  44847.24   \\
 2.3 & 18718.39$\pm$0.37  & 18700.28$\pm$0.93  &  20159.33$\pm$0.64 &    18728.15     &  18716.36   \\
 2.8 & 6021.23$\pm$0.16   & 6015.01$\pm$0.45   &   5575.28$\pm$5.37 &      6021.75    &   6020.67   \\
 3.2 & 2334.11$\pm$0.11   & 2331.66$\pm$0.37   &   2271.94$\pm$1.45 &      2333.37    &   2333.58   \\
 3.8 & 508.75$\pm$0.08    & 507.53$\pm$0.23    &    537.78$\pm$1.77 &       508.66    &    508.42   \\
 4.3 & 118.019$\pm$0.022  & 117.879$\pm$0.059   &    102.24$\pm$0.32 &       118.13    &    117.93   \\
 4.5 & 58.475$\pm$0.013   & 58.425$\pm$0.034   &     49.76$\pm$0.29 &       58.547    &     58.406  \\
 4.7 & 24.229$\pm$0.012   & 24.205$\pm$0.031   &     19.49$\pm$0.27 &       24.312    &     24.220  \\
 5.6 & -11.003$\pm$0.005  & -11.014$\pm$0.015  &    -10.89$\pm$0.17 &      -10.9957   &    -11.0048 \\
 5.9 & -10.184$\pm$0.005  & -10.195$\pm$0.014  &    -10.31$\pm$0.15 &      -10.1793   &    -10.1865 \\
 7.4 & -3.332$\pm$0.004   & -3.335$\pm$0.012   &     -2.57$\pm$0.18 &       -3.3297   &     -3.3330 \\
\end{tabular}
\label{}
\end{ruledtabular}
\end{table}

\end{document}